\documentclass[conference,letter]{IEEEtran}
\usepackage{subfigure}
\usepackage{graphicx}
\usepackage{latexsym}
\usepackage{amsxtra} 
\usepackage{amsmath}
\usepackage{amsfonts}
\usepackage{marvosym}
\usepackage{scalefnt}
\usepackage{color,tabularx}
\usepackage{makecell}
\usepackage{booktabs,multicol,multirow}
\usepackage{algorithm2e}
\newcommand{\href}[2]{#2}

\definecolor{mygray}{gray}{0.6}
\newcommand{\tti}[2]{\textcolor{mygray}{#1} + #2 s}
\newcommand{\mainResult}[1]{\textbf{#1}} 

\usepackage{fancyhdr,lipsum}
\fancypagestyle{firstpage}{
  \fancyhf{}
  \fancyhead[C]{To appear at the 23rd Design, Automation and Test in Europe (DATE 2020). Grenoble, France.}
  \fancyfoot[C]{\thepage}
}

\begin{document}\bstctlcite{IEEEexample:BSTcontrol}
\title{TFApprox: Towards a Fast Emulation of DNN Approximate Hardware Accelerators on GPU\vspace*{-0.5em}}

\author{
\IEEEauthorblockN{Filip Vaverka, Vojtech Mrazek, Zdenek Vasicek, Lukas Sekanina}
\IEEEauthorblockA{\scalefont{0.95}{Brno University of Technology, Faculty of Information Technology, IT4Innovations Centre of Excellence, Brno, Czech Republic}\\
Email: \{vaverka, mrazek, vasicek, sekanina\}@fit.vutbr.cz}
}



\maketitle

\thispagestyle{firstpage}

\begin{abstract}
Energy efficiency of hardware accelerators of deep neural networks (DNN) can be improved by introducing approximate arithmetic circuits. In order to quantify the error introduced by using these circuits and avoid the expensive hardware prototyping, a software emulator of the DNN accelerator is usually executed on CPU or GPU. However, this emulation is typically two or three orders 
of magnitude slower than a software DNN implementation running on CPU or GPU and operating with standard floating point arithmetic instructions and common DNN libraries. The reason is that there is no hardware support for approximate arithmetic operations on common CPUs and GPUs and these operations have to be expensively emulated. In order to address this issue, we propose an efficient emulation method for approximate circuits utilized in a given DNN accelerator which is emulated on GPU. All relevant approximate circuits are implemented as look-up tables and accessed through a texture memory mechanism of CUDA capable GPUs. We exploit the fact that the texture memory is optimized for irregular read-only access and in some GPU architectures is even implemented as a dedicated cache. This technique allowed us to reduce the inference time of the emulated DNN accelerator approximately 200 times with respect to an optimized CPU version on complex DNNs such as ResNet.
The proposed approach extends the TensorFlow library and is available online at \href{https://github.com/ehw-fit/tf-approximate}{github.com/ehw-fit/tf-approximate}.
\end{abstract}

\section{Introduction}
As the training of \emph{deep neural networks} (DNNs) is a very computationally expensive procedure for state-of-the-art DNNs (containing hundreds 
of layers and ten millions of parameters), it is usually performed on clusters of graphic processing units (GPUs) or even on supercomputers.  If a more energy-efficient implementation is required, a specialized \emph{hardware DNN accelerator} is developed, for example, see Google`s TPU~\cite{Jouppi:2017} or various academia neural chips~\cite{chen2014dadiannao,lu2017flexflow}. Such accelerators typically implement only the inference procedure for fully trained DNNs. In addition to simplifying the DNN architecture (by the so-called pruning), a significant power consumption reduction of the DNN hardware accelerator can be obtained by introducing approximate memory and \emph{approximate arithmetic circuits}~\cite{Mittal:2016}. As applications of DNNs are highly error-resilient~\cite{Venkataramani:axnn,Sarwar:2018}, the error introduced by approximate (i.e. simplified) implementations of circuits and programs often remains acceptable and, in most cases, invisible to the user. It is, however, important to quantify the error introduced by employing approximate circuits and find the best tradeoff between the error and power requirements prior a real hardware design is started. This is usually done with the help of software platforms developed for DNN design and training (such as TensorFlow~\cite{tensorflow2015-whitepaper}). These platforms are, however, optimized for standard floating point arithmetic operations available in processors and GPUs and use standard libraries of mathematical functions. If these operations are replaced with user-specific (fixed point) operations (e.g., approximate multipliers), the DNN execution is slowed down by several orders of magnitude on processors as well as GPUs. The reason is that there is no hardware support for approximate arithmetic operations on common processors and GPUs and these operations have to be expensively emulated. 

In order to analyze the DNN error after introducing approximate arithmetic operations into the computation datapath of a hardware DNN accelerator, we propose a new GPU-based emulation platform for DNN accelerators containing approximate circuits. The reason for choosing a GPU is that (i) complex DNNs are usually trained using software tools such as TensorFlow that are highly optimized for GPUs and (ii) GPUs provide higher performance than common processors for DNN applications. Note that determining a suitable approximate implementation of arithmetic operation for a given DNN and a given application requires evaluating many candidate approximate operations and, in most cases, performing additional parameter fine-tuning (i.e. re-training). Existing DNN development platforms do not support the aforementioned approach. For example, Ristretto (operating over Caffe) is capable of evaluating various number representations that can be employed in DNNs and find an optimum bit width for these operations~\cite{Ristretto}. However, it does not support approximate arithmetic operations, i.e. only common operations with reduced bit widths can be executed. 

In the proposed emulation platform, all relevant approximate circuits are implemented as look-up tables and accessed through a texture memory mechanism of CUDA capable GPUs. We exploit the fact that the texture memory is optimized for irregular read-only access and in some GPU architectures is even implemented as a dedicated cache. This technique allowed us to reduce the inference time of the emulated DNN accelerator approximately 200 times with respect to an optimized CPU version on complex DNNs such as ResNet.
The proposed approach has been embedded into the TensorFlow library. 

\section{CNNs with Approximate Multipliers Emulated on GPU}
Various quantization schemes have been applied within DNNs~\cite{fengfu:2016,bnn:2016,Benoit:quantization}.
The affine transformation represents the most preferred technique which is employed, for example, also in TensorFlow (TF).
It allows an efficient implementation of arithmetic operations using only integer arithmetic on the quantized values.
This quantization scheme $Q:\mathbb{R}\rightarrow \mathbb{N}$ maps a real number $r$ to an integer $i$ to ensure the following equality:
\begin{equation}\label{eq:affine}
r = \alpha(i - \beta),
\end{equation}
where $\alpha$ and $\beta$ are two constants corresponding with the so-called \emph{scale} and \emph{zero-point}~\cite{Benoit:quantization}.
While $\alpha$ is a positive real number, $\beta$ should be a number of the same type as $i$.
The constants are chosen in such a way that the real value $r = 0$ is exactly representable by a quantized value. 
This requirement is important because many computations result in 0 and it is highly undesirable to propagate a non-zero quantization error to next layers. In addition, zero padding is often applied.

To improve the efficiency, the convolution on two functions is typically implemented using the $N\times N$ matrix multiplication. 
Given two input matrices denoted as $input$ and $filter$, both having $N\times N$ real elements, the output value corresponding with an element of the matrix product at position $[i,j]$ is calculated as follows:
\begin{equation}\label{eq:conv}
output[i,j] = \sum_{k=1}^{N}  input[i,k] filter[k,j]
\end{equation}
When we apply the affine transform (Eq.~\ref{eq:affine}), the output is: 
\begin{equation}
output[i,j] = \sum_{k=1}^{N}  \alpha_1(\overline{input}[i,k] - \beta_1) \alpha_2(\overline{filter}[k,j] - \beta_2)
\end{equation}
where $\overline{x}$ represents the quantized value of $x$. This expression can be rewritten as
\begin{multline}\label{eq:c1}
output[i,j] = \alpha_1\alpha_2 \sum_{k=1}^{N} \overline{input}[i,k] \overline{filter}[k,j] - \\
\alpha_2\beta_2 \sum_{k=1}^{N} input[i,k] - 
\alpha_1\beta_1 \sum_{k=1}^{N} filter[k,j] + 
 N\alpha_1\alpha_2\beta_1\beta_2
\end{multline}

The first sum in Eq.~\ref{eq:c1} represents the summation on the quantized values which can be calculated using the integer operations. In hardware, this can efficiently be implemented using an integer MAC circuit. The remaining two sums can also be done on the quantized values but it is beneficial for our purposes to express them in terms of the real numbers. The obtained equation describes 
the convolution on two independently quantized inputs followed by the dequantization.

As analyzed in~\cite{Jouppi:2017}, the 8-bit operations are sufficient for DNN accelerators based on integer arithmetic. It means that we need a MAC unit consisting of an 8-bit multiplier and 32-bit accumulator to calculate a single element of the matrix product~\cite{Benoit:quantization}.
To emulate the behavior of the approximate multipliers employed in the MAC, an 8-bit approximate multiplier is used for determining the product $\overline{input}[i,k] \overline{filter}[k,j]$ in our DNN approximate hardware accelerator.

Several types of 2D convolutional layers that implement a variant of Eq.~\ref{eq:conv} are available in TF.
To support the approximate multiplications in the training as well as inference process seamlessly without the necessity to rewrite the training algorithms already implemented in TF, we propose to introduce an alternative approximate 2D convolutional layer to each type of the 2D convolution implementing a variant of Eq.~\ref{eq:c1}. 
The approximate layer reads two floating-point inputs and produces a single floating-point output which has the same range as if we use the original convolutional layer.

Compared to the common convolutional layers, we need to provide some additional information: four scalars specifying the quantization coefficients, a model of the approximate multiplier, expected range of the quantized values ([-128, 127] for signed, [0, 255] for unsigned multipliers) and requested round mode for the rounding applied during the quantization. In fact, the coefficients can be calculated independently for each input vector using the knowledge of the range of the inputs (i.e. minimum and maximum value). The approximate multiplication is specified by means of its truth table. This approach offers the highest throughput and does not cause any limitations as the truth table for an 8-bit multiplier occupies only 128~kB. 

\begin{figure}[b]%
    \centering\vspace{-1.5em}
    \includegraphics[width=.75\columnwidth]{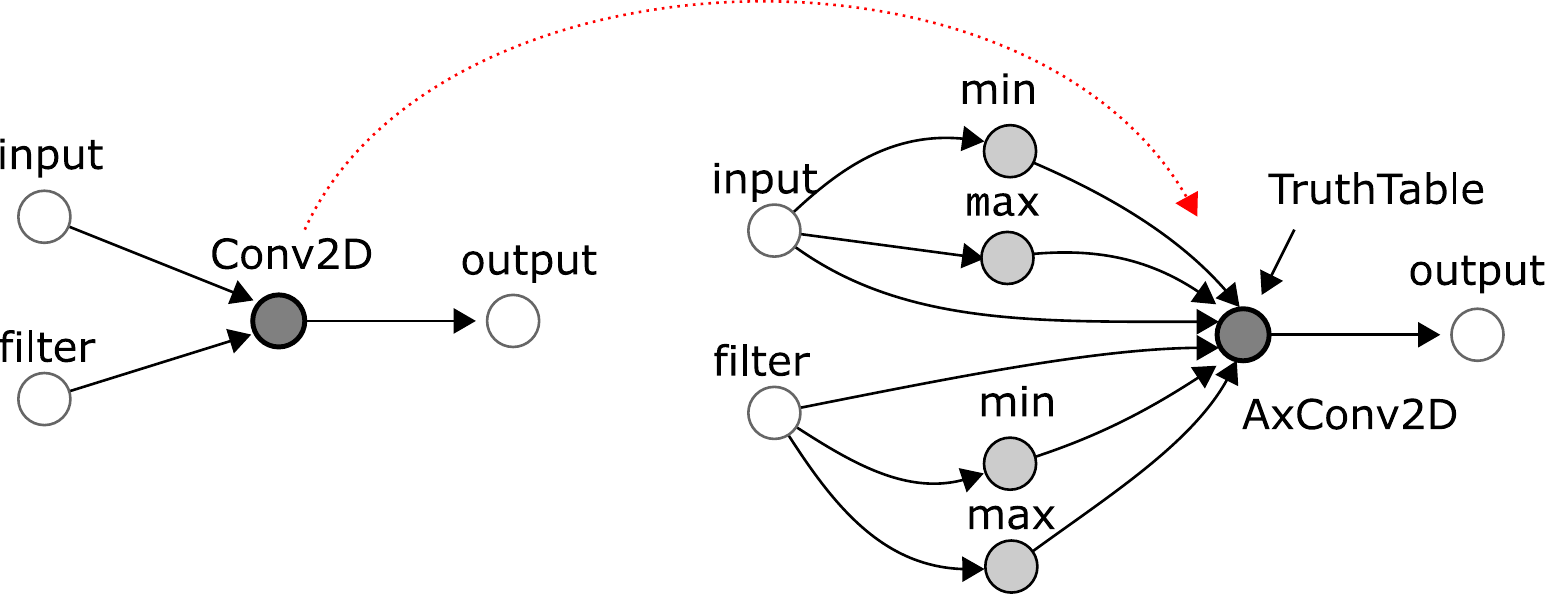}
    \caption{Introducing the approximate convolutional layer (AxConv2D) into the existing graph consisting of a single convolutional layer Conv2D.}
    \label{fig:graph}
\end{figure}

The design flow is as follows. Firstly, a DNN model is created or loaded in TF. Then, all convolutional layers are identified and replaced by corresponding approximate variants. During this process, the minimum and maximum operators are inserted into the computational path and connected to the approximate layers. At the end, we obtain a transformed graph which is suitable for the inference as well as training because the minimum and maximum values of the input tensors are determined once per a batch.
A part of the original and transformed graph is shown in Fig.~\ref{fig:graph}.

\section{GPU implementation}
The 2D convolution operation in TF typically expects two 4D input tensors and produces another 4D tensor provided that the stride and dilation parameters are specified. The first input tensor represents a batch of 3D input images given in NHWC format (Batch$\times$Height$\times$Width$\times$Channels), where the number of channels corresponds with the fastest changing index. The second tensor is a set of 3D filters (or kernels of the convolution) stored in the Height$\times$Width$\times$Channels$\times$Count format, where Count specifies the number of filters applied to the same input. The output of the convolution shares the same layout as the input data; however, the height and width are determined according to the shape of the kernel and the depth of each output image depends on the number of applied filters. The approximate version of the 2D convolution is extended by four scalar inputs that provide the minimum and maximum values computed independently for each input vector. 

In~\cite{iccad19prelim}, the authors applied a direct approach to implement the TF-compatible approximate 2D convolution. Unfortunately, only a CPU platform was supported. The method  directly stems from the definition of the convolution operation and the format of inputs. This leads to a system of nested loops (over each input image in the batch, each output pixel, each output channel etc.) which is difficult to efficiently parallelize on GPUs. To aviod this issue, we selected the General Matrix-matrix multiplication (GEMM) approach for our CUDA-based GPU implementation. Similarly to the CPU-based implementation, we adopted the idea to implement the inner multiplication between each input and filter values using a lookup table.

\begin{algorithm}[b]\vspace{-1em}%
    \SetKwFunction{SplitData}{SplitData}
    \SetKwFunction{ComputeDeqCoeffs}{ComputeCoeffs}
    \SetKwFunction{ChunkToCols}{Im2Cols}
    \SetKwFunction{ApproxGEMM}{ApproxGEMM}
    \SetKwFunction{AppendOutChunk}{AppendOutput}
    \SetKwInOut{Parameters}{Parameters}
    
    \KwIn{input\_batch, filters, input\_batch\_range, filters\_range}
    \Parameters{strides, dilations, lookup\_table}
    \KwOut{output\_batch}
    $\alpha_1, \beta_1$ $\leftarrow$ \ComputeDeqCoeffs{input\_batch\_range}\;
    $\alpha_2, \beta_2$ $\leftarrow$ \ComputeDeqCoeffs{filters\_range}\;
    $\mathbf{S_f}$ $\leftarrow$ $\sum_{h,w,d} filters[h,w,d,c]$\;
    chunks $\leftarrow$ \SplitData{input\_batch, chunk\_size}\;
    \ForEach{chunk $\in$ chunks}{
        $\mathbf{\overline{M_p}}, \mathbf{S_p}$ $\leftarrow$ \ChunkToCols{chunk, $\alpha_1$, $\beta_1$, strides, dilations}\;
        $\mathbf{O}$ $\leftarrow$ \ApproxGEMM{$\mathbf{\overline{M_p}}$, $\mathbf{S_p}$, filters, $\mathbf{S_f}$, $\alpha_1$, $\beta_1$, $\alpha_2$, $\beta_2$, lookup\_table}\;
        output\_batch $\leftarrow$ \AppendOutChunk{output\_batch, $\mathbf{O}$}
    }
    \caption{Approximate 2D convolution}%
    \label{alg:approx_2d_conv}%
\end{algorithm}

The GEMM-based approach splits the convolution into two separate operations. (i) The patch matrix in which each row corresponds to a single position of the kernel is constructed (the image-to-columns phase). (ii) The patch matrix is multiplied with the filters matrix in which each column corresponds to a single filter (GEMM phase). The GPU implementation of the approximate 2D convolution mostly follows this structure, but extends it with a few auxiliary computations to precompute the constant terms of Eq.~\ref{eq:c1}. 

Algorithm~\ref{alg:approx_2d_conv} describes the high-level structure of our implementation. At the beginning, the quantization parameters $\alpha$, $\beta$ are computed using the input range information (the minimum and maximum values provided separately for each input). Then, the filter-only sum $\mathbf{S_f}$ corresponding with the third sum in Eq.~\ref{eq:c1} is computed. The input batch is then split into chunks of a constant size to decouple memory usage from convolution parameters. Next, each chunk is converted to a matrix of 8-bit integer values $\mathbf{\overline{M_p}}$, in which each row (patch) corresponds to single position of the convolution kernel. At the same time, the dequantization sum for each patch is also computed and stored as a vector $\mathbf{S_p}$. Finally, the matrix $\mathbf{\overline{M_p}}$ is multiplied by the matrix of filters (which are quantized at the same time) and the results are dequantized using pre-computed correction terms. The dequantized result of matrix multiplication is appended to the output of the operation.

\subsection*{(i) Image-to-Columns phase}
The image-to-columns phase (\textit{Im2Cols} function in Algorithm \ref{alg:approx_2d_conv}) can easily be parallelized in CUDA by running a single thread for each output value of $\mathbf{\overline{M_p}}$. However, in order to compute the sums in vector $\mathbf{S_p}$ in a single pass over the data, one may limit the number of parallel threads to a single thread per patch or the thread block size has to be tied to the patch length (the reduction can thus be performed in a shared memory). As these approaches can limit the level of parallelism or flexibility of the solution, we opted for a slightly different way to compute these sums. 

The thread block size in our solution is fixed and independent of the patch length. This means that any given thread block can process
one or even several patches at the time. Multiple reductions over the values processed by consecutive threads have to be performed and each result has to be added to an appropriate element of $\mathbf{S_p}$. In CUDA, this can be done (with a reasonable efficiency) by loading input values into the shared memory and performing a prefix scan which allows extracting the partial sums at the end of each patch. These results are then added atomically (using \textit{atomicAdd}) to the appropriate element of $\mathbf{S_p}$ as the rest of the patch may be processed by other thread blocks.

\begin{table*}
    \caption{Time required to process CIFAR-10 dataset on CPU and GPU when accurate and approximate convolutional layers are employed.}%
    \vspace*{-1em}
\centering%
\resizebox{.99\textwidth}{!}{
\begin{tabular}{l|rr|rr|rr|rr|rr}
\toprule 
& \multicolumn{2}{c|}{DNN parameters} & \multicolumn{2}{c|}{Accurate Conv2D} & \multicolumn{2}{c|}{Approximate AxConv2D} &  \multicolumn{2}{c|}{Approx. overhead} &  \multicolumn{2}{c}{Speedup GPU vs CPU}\\
\bf DNN & \boldmath$L$ & \bf \# MACs & \bf CPU & \bf GPU & \bf CPU & \bf GPU & \bf CPU & \bf GPU & \bf Accurate & \bf Approximate \\
\midrule
ResNet-8 &   7 & $ 21\times10^6$ & \tti{0.2}{4.4}&  \tti{1.8}{0.2} & \tti{0.2}{341} & \tti{1.7}{1.5} & 337 s  & 1.2 s & 2.3 $\times$  & \mainResult{106.8 $\times$}  \\
ResNet-14 &  13 & $ 35\times10^6$ & \tti{0.2}{7.4}&  \tti{1.9}{0.3} & \tti{0.2}{724} & \tti{1.8}{3.1} & 718 s  & 2.7 s & 3.5 $\times$  & \mainResult{148.8 $\times$}  \\
ResNet-20 &  19 & $ 49\times10^6$ & \tti{0.2}{10.4}&  \tti{1.8}{0.5} & \tti{0.2}{1105} & \tti{1.8}{4.7} & 1096 s  & 4.3 s & 4.7 $\times$  & \mainResult{170.2 $\times$}  \\
ResNet-26 &  25 & $ 63\times10^6$ & \tti{0.2}{13.4}&  \tti{1.9}{0.6} & \tti{0.2}{1489} & \tti{1.8}{6.2} & 1477 s  & 5.6 s & 5.5 $\times$  & \mainResult{185.0 $\times$}  \\
ResNet-32 &  31 & $ 77\times10^6$ & \tti{0.3}{16.3}&  \tti{1.9}{0.7} & \tti{0.3}{1876} & \tti{1.9}{7.9} & 1861 s  & 7.3 s & 6.5 $\times$  & \mainResult{191.0 $\times$}  \\
ResNet-38 &  37 & $ 91\times10^6$ & \tti{0.3}{19.3}&  \tti{1.9}{0.8} & \tti{0.3}{2259} & \tti{1.9}{9.4} & 2241 s  & 8.6 s & 7.3 $\times$  & \mainResult{200.1 $\times$}  \\
ResNet-44 &  43 & $106\times10^6$ & \tti{0.3}{22.3}&  \tti{1.9}{0.9} & \tti{0.3}{2640} & \tti{2.0}{10.9} & 2620 s  & 10.0 s & 8.0 $\times$  & \mainResult{205.6 $\times$}  \\
ResNet-50 &  49 & $120\times10^6$ & \tti{0.3}{25.2}&  \tti{1.9}{1.1} & \tti{0.3}{3025} & \tti{2.0}{12.6} & 3003 s  & 11.7 s & 8.6 $\times$  & \mainResult{207.2 $\times$}  \\
ResNet-56 &  55 & $134\times10^6$ & \tti{0.3}{28.1}&  \tti{1.9}{1.2} & \tti{0.3}{3409} & \tti{2.0}{13.9} & 3384 s  & 12.8 s & 9.2 $\times$  & \mainResult{214.4 $\times$}  \\
ResNet-62 &  61 & $148\times10^6$ & \tti{0.3}{31.1}&  \tti{1.9}{1.3} & \tti{0.3}{3796} & \tti{2.3}{15.5} & 3767 s  & 14.7 s & 10.0 $\times$  & \mainResult{213.2 $\times$}  \\
\bottomrule
\end{tabular}}%
\label{tab:res}%
\vspace*{-1em}
\end{table*}

\subsection*{(ii) Matrix multiplication phase}
The matrix multiplication phase (\textit{ApproxGEMM} function in Algorithm \ref{alg:approx_2d_conv}) is implemented as a typical tiled GEMM, in which the threads of the block have to load a 2D tile from each matrix into the shared memory and each thread computes a single output value. The tiles in the shared memory are quantized and stored as \textit{uint} to avoid possible shared memory access conflicts. The multiplication of quantized 8-bit values is implemented by a lookup table containing $256^2$ 16-bit values stored in GPU memory and cached in L1 or L1 texture cache. To manage this in CUDA, \textit{cudaTextureObject\_t} is used to store the table and \textit{tex1Dfetch\textless ushort\textgreater} to perform the lookup based on the index created by stitching the multiplied 8-bit values into a single 16-bit value. The results of multiplication (lookup) operations are accumulated in a 32-bit floating point accumulator. The last step is to perform dequantization
and a correction according to Eq.~\ref{eq:c1} using $\alpha_i$, $\beta_i$ terms and precomputed constants stored in vectors $\mathbf{S_p}$ and $\mathbf{S_f}$.

\section{Results}

The proposed emulation platform was implemented in C++ using NVIDIA CUDA Toolkit 10.1 and integrated into TensorFlow library. Its performance was evaluated on a residual ResNet~\cite{resnet:2015} DNN because it enabled us to easily configure the number of building blocks and thus the number of 2D convolutional layers $L$ and MAC operations (see Tab.~\ref{tab:res}). We focused on the performance evaluation of the approximate layers. Note that the accuracy is the same as if we use the quantization followed by dequantization available in TensorFlow. 
The experiments were conducted on Intel Xeon E5-2620 CPU and NVIDIA GTX 1080 CPU. We used CIFAR-10 dataset containing $10^4$ input images having $32\times32\times3$ pixels each. Ten pre-trained ResNet models were used whose parameters are provided in Tab.~\ref{tab:res}. Only the inference process is considered to avoid any bias. 
The content of the LUT table implementing an approximate multiplier does not have any impact on the execution time. The evaluation of the data set is divided in 10 batches consisting of 1000 images each.

\begin{figure}[b]
    \centering\vspace{-2em}
    \includegraphics[width=\linewidth]{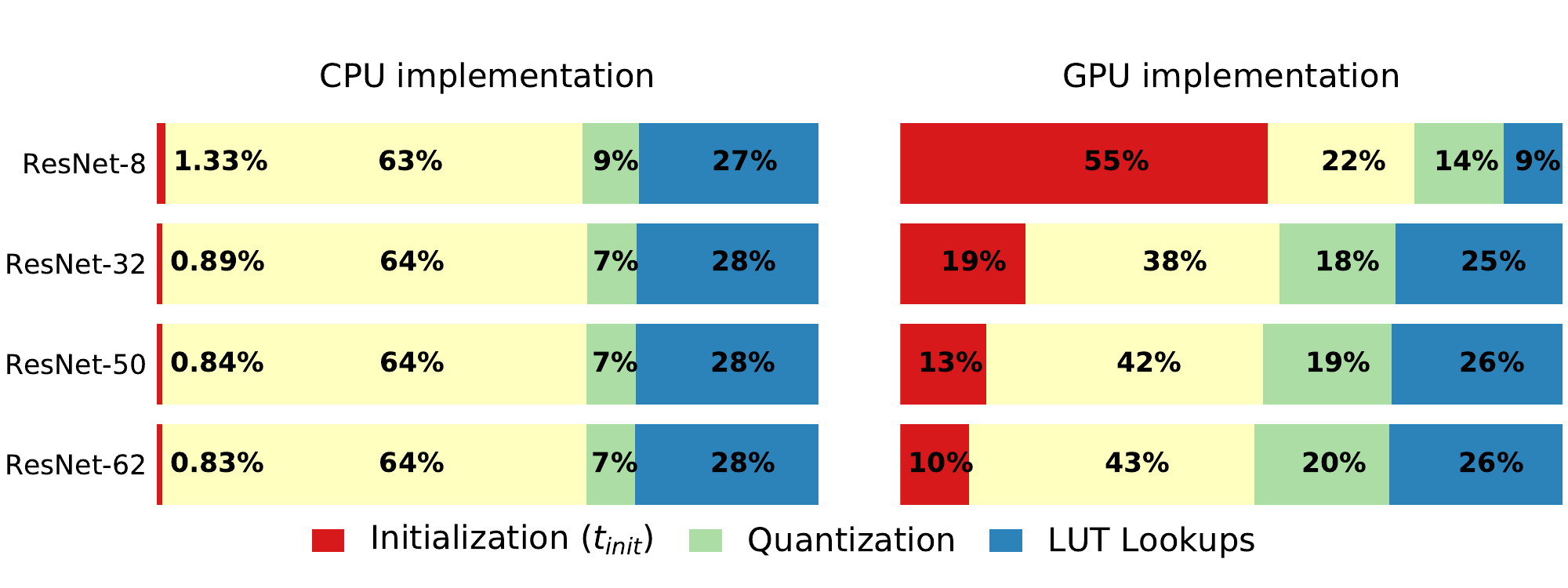}\vspace{-1em}
    \caption{Distribution of the total computational time $t_{init}+t_{comp}$.}
    \label{fig:prop}
\end{figure}

Our implementation is compared with the native and highly optimized implementation for CPUs and GPUs already available in TF (columns `accurate Conv2D`) as well as with the CPU-based approach (first column in `approximate Conv2D`).
For each DNN, Tab.~\ref{tab:res} reports the time in form of $t_{init} + t_{comp}$, where $t_{init}$ is the initialization of the computation (including the memory allocation and data transfer which is critical especially in case of GPUs) and $t_{comp}$ required to process the whole data set by the DNN. While $t_{init}$ is nearly constant (the same data set is used in all cases), $t_{comp}$ increases linearly with increasing the number of MACs. The last two columns contain the achieved speedup on GPU which also grows linearly. The proposed accelerator achieves significantly better performance compared to the CPU-based one (see the last column). The overhead introduced by the necessity to perform quantization, LUT lookup and dequantization is much smaller. For ResNet-62, the computation time was reduced from 3796~s to 15.5~s. The overhead due to the emulation of the approximate computations is still quite high (14.7~s) but the total time is now acceptable for the practical usage. 
Fig.~\ref{fig:prop} shows a more detailed analysis of the total computation time for some configurations. For ResNet-62, 26\% of the total time is caused by the LUT lookups, 20\% is due to the quantization, dequantization and computation of min/max, 10\% is spent in the initialization phase and the rest is the remaining computation (Im2Cols, GEMM, etc.).

%
%

\section{Conclusions}
 We proposed an efficient emulation method for DNN accelerators containing approximate multipliers. This method allowed us to reduce the inference time of the emulated DNN accelerator approximately 200 times with respect to an optimized CPU version on complex DNNs such as ResNet. This opens new ways to automated design of approximate DNN accelerators in which many candidate designs have to be quickly evaluated.

\section*{Acknowledgements}
\emph{This work was supported by Czech Science Foundation project 19-10137S.}

\bibliographystyle{IEEEtran}
\bibliography{IEEEabrv,date20}

\end{document}